\documentclass{article}
\usepackage[letterpaper, margin=0.71in]{geometry}
\usepackage{multicol}
\usepackage{amsmath}
\usepackage{graphicx}
\usepackage{caption}
\usepackage{subcaption}
\usepackage{float}
\usepackage{enumitem}
\usepackage{xcolor}
\usepackage{soul}
\usepackage[font=small,labelfont=bf]{caption}
\usepackage{chemformula}
\usepackage{hyperref}
\usepackage{gensymb}
\usepackage{graphicx} 
\title{Magnetism at the interface of \ch{MoSe2/V2O5} heterostructures}
\author{Rohin Sharma\textsuperscript{1}, 
Diem Thi-Xuan Dang\textsuperscript{1}, 
Lilia M. Woods\textsuperscript{1,*}}
\date{}
\begin{document}
\maketitle
\noindent
\begin{center}
\emph{\textsuperscript{1} University of South Florida, Department of Physics, Tampa, FL 33620, USA.}\\
\noindent
\\
* Corresponding email: lmwoods@usf.edu
\paragraph{}
\end{center}
\paragraph{}
\hrule
\vspace{5mm}
\textbf{Abstract}: Magnetism in doped transition metal dichalcogenide monolayers and van der Waals interfaced materials have motivated the search for sustainable magnetic states at the nanoscale with the prospect of building devices for spintronics applications. In this study, we report the existence of magnetism in a heterostructure made up of an \ch{MoSe2} transition metal dichalcogenide monolayer and a \ch{V2O5} substrate. Using density functional theory simulations, we find that ferromagnetic ordering can be found in the \ch{MoSe2/V2O5} heterostructure even though the individual components are nonmagnetic. By examining the electronic structure and magnetic properties of this system we find how the occurring ferromagnetism evolves if the transition metal dichalcogenide or the \ch{V2O5} substrate can host point defects. Our study suggests that the balance between charge transfer and spin reorganization can lead to interface magnetism in novel hybrid materials.
\vspace{5mm}
\hrule
\begin{multicols}{2}
\section*{Introduction}
Two-dimensional materials can host an array of extraordinary properties. A striking example is the existence of 2D magnetism in different atomically thin crystals~\cite{gibertini2019magnetic}. Although many nanoscaled systems are metallic, ferromagnetism in transition metal dichalcogenides (TMDs) substitutionally doped by V, Fe, Cr or other transition metal atoms has opened up the field of 2D dilute magnetic semiconductors~\cite{yun2020ferromagnetic, pham2020tunable, tiwari2021magnetic}. The tunable ferromagnetic (FM) state in doped TMDs is robust at room temperature and it can be controlled not only by magnetic fields, but also by electric fields and light \cite{lee2021role, ortiz2021light, ortiz2024transition}. Due to their weak van der Waals coupling, TMDs can also be stacked with other materials forming heterostructures with properties distinct from their constituents~\cite{hung2023enhanced, wang2015spin, thi2022enhanced, lee2020enhanced, phan2021perspective}. The high-quality interfaces and integrated opto-electro-magnetic properties have shown that such heterostructures maybe suitable for novel magnetic and spintronic devices at the nanoscale \cite{xie2021chemistry, zhang2016large}. For example, experiments have shown that thermally generated spin currents in ferromagnetic-TMD-heavy metal heterostructures are possible due to an enhanced spin Seebeck coefficient~\cite{dastgeer2019distinct, lee2020enhanced}. Spin filtering across interfaces and gate-tunable magnetic ordering in TMD heterostructures have  been demonstrated as well~\cite{wang2015spin, zatko2019band, nguyen2021gate}.

Interface magnetism has also been observed in heterostructures composed of nonmagnetic components. In particular, the polar discontinuities between thin films of insulating \ch{LaAlO3} and \ch{SrTiO3} oxides can form a conducting interface~\cite{ohtomo2002artificial, ohtomo2004high} with electronic density larger by an order of magnitude compared to the density of other typical semiconductor interfaces. The \ch{LaAlO3/SrTiO3} heterostructure can experience ferromagnetism, superconductivity and strong spin-orbit coupling~\cite{bert2011direct, lee2013titanium, ben2010tuning}. First principles calculations have shown that the reconstruction at the interface and the interplay between geometrically confined carrier spins and strong correlation effects give rise to a FM state~\cite{bi2014room, janicka2008magnetism, pentcheva2006charge}. Similar to the TMDs with tunable magnetism via the doping concentration, the \ch{LaAlO3/SrTiO3} ferromagnetism can also be controlled by the carrier concentration of the conductive layer at the interface \cite{bi2014room}. \ch{LaAlO3/SrTiO3} paves the way for optically transparent oxide spintronics with devices that can be integrated in optoelectronic applications~\cite{kaur2024room}.

The quest for sustainable magnetism at the nanoscale motivates searching for new ways and materials, that can realize robust and controllable phenomena. In this paper, we show that ferromagnetism can exist in a heterostructure formed by a \ch{MoSe2} monolayer and \ch{V2O5} substrate. This van der Waals system is composed of two nonmagnetic components as one of them is a TMD monolayer while the other one is an oxide semiconductor. The magnetic response is confined at the atomically sharp interface, and it resembles a cross-over between the 2D dilute magnetic semiconductors and the oxide heterostructures discussed above. 
\end{multicols}
\begin{figure}
    \centering
    \includegraphics[width=0.9\linewidth]{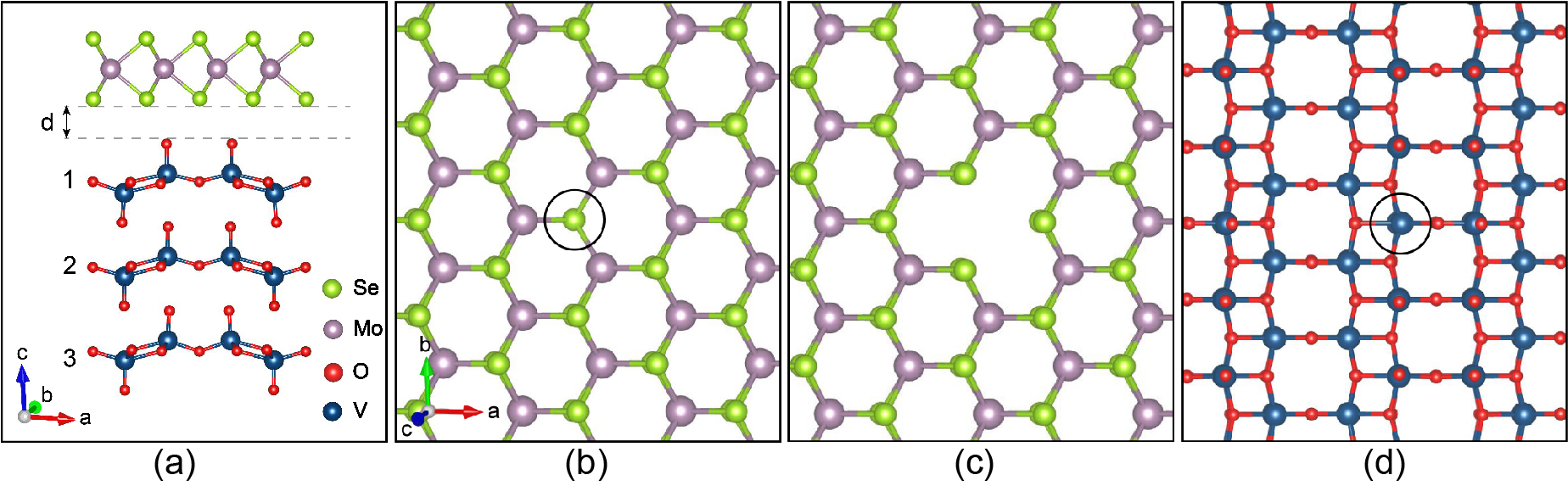}
    \caption{(a) Side view of the \ch{MoSe2/V2O5} heterostructure with monolayer \ch{MoSe2} separated by a distance $d$ from the trilayer \ch{V2O5} substrate; (b) Monolayer \ch{MoSe2} with a point defect of a Se vacancy -\ch{V_{Se}}; (c) Monolayer \ch{MoSe2} with a point defect of a Mo vacancy -\ch{V_{Mo}}; (d) Top layer (labeled as (1)) of the \ch{V2O5} substrate with a point defect of an O vacancy -\ch{V_{O}}.}
    \label{fig:structure}
    \vspace{1mm}
    \hrule  
\end{figure}
\begin{multicols}{2}
\section*{Computational Methods}
Our calculations are based on the DFT + U approach using Predew-Burke-Ernzerhof (PBE) formulation~\cite{perdew1996generalized} of the Generalized Gradient Approximation (GGA) implemented in plane-wave pseudopotential code of Vienna Ab-initio Simulation Package (VASP)~\cite{kresse1993ab, kresse1996efficient}. The valence electrons are approximated with the Projector Augmented Wave (PAW) method~\cite{kresse1999ultrasoft} with a plane-wave cut-off of 450~eV. The strong correlation effects  due to on-site Coulomb are accounted for by using U=3.4~eV found to be appropriate  to describe the correlation effects due to localized $d$ orbitals of the V atoms~\cite{sihi2020detailed}. To describe the inter-layer and interface environment dominated by vdW forces, we have considered the long-range dispersion energy corrections in the semi-empirical approach proposed by Grimme (DFT-D3)~\cite{grimme2011effect}. We have also studied our system with the Spin-orbit coupling included to point out major differences due to this relativistic effect.

The construction of the heterostructures supercells is done by applying appropriate transformation on the standalone components. In order to make the rhombohedral MoSe$_2$ structure compatible with the orthogonal V$_2$O$_5$ substrate, the TMD cell is obtained by first applying $\sqrt{3}\times 1$ transformation on the unit cell, followed by a $2\times 3$ transformation. The cell of V$_2$O$_5$ is obtained by applying $2\times 3$ transformation on its unit cell. The atomic registry of the heterostructures is taken to ensure short starting distance between the defect and the atoms from the other component to maximize reactivity. Also, in forming the heterostructure, the MoSe$_2$ lattice is slightly stretched while the V$_2$O$_5$ is slightly squeezed, a usual practice when simulating incommensurate lattice structures. To eliminate the interaction due to periodicity on the c axis, a vacuum layer of 17~$\AA$ is added between the heterostructure layers. Upon relaxation, we find the $a,b$ lattice parameters compared to their standalone counterparts have stretched by 1\% and 6.7\% for the TMD layer and they have compressed by -0.2\% and -2.5\% for the substrate.

The relaxation criteria of all systems are used by allowing all cell parameters to change with $10^{-5}$~eV energy and 0.01~eV/$\AA$ force relaxation criteria. After examining the k-point sampling of the Brillouin zone, structural relaxation and single point calculations are performed with a $3\times3\times1$ $\Gamma$--centered Monkhorst--Pack k-point mesh. To overcome the zone folding problem occurring in simulations involving supercells, the band structures are unfolded and projected on the first Brillouin zone of the TMD ($\Gamma-M-K-\Gamma$). The $k$ vector mesh along the high--symmetry directions in the primitive BZ is generated by adopting the vaspkit code~\cite{wang2021vaspkit}. The visualization of atomic position, charge density difference and spin re-distribution is obtained using the  VESTA package~\cite{momma2011vesta}.
\section*{Results and discussion}
\subsection*{Structural properties}
We consider a system composed of a \ch{MoSe2} monolayer with $P6_3$/mmc symmetry on top of a \ch{V2O5} substrate, as shown in Fig.~\ref{fig:structure} (a). The constructed supercell contains 36 atoms from the TMD monolayer. \ch{V2O5} has several polytypes, but here we take the $\alpha$--phase, the most common and thermodynamically stable one~\cite{hu2023vanadium, ronquillo2016synthesis}. The $\alpha$--\ch{V2O5} is a layered material with an orthorhombic structure in the $Pmmn$ space group. For the heterostructure, the $\alpha$--\ch{V2O5} substrate, separated by a distance $d$ from \ch{MoSe2}, is modeled by three layers with 126 atoms in the supercell. Recent studies have shown that indeed thin films of \ch{V2O5} have been synthesized and the electronic structure of very thin substrates is very similar to the one for bulk~\cite{scanlon2008ab, das2019structural, nushin2024preparation}. Thus, the constructed heterostructure may serve as the prototype of a TMD monolayer above the bulk \ch{V2O5} substrate. 

Several variations of the \ch{MoSe2/V2O5} system are also simulated by taking different types of point defects. In Fig.~\ref{fig:structure}~(b, c), the top view of a Se~(-\ch{V_{Se}}) and Mo~(-\ch{V_{Mo}}) vacancies in the \ch{MoSe2} monolayer are shown. Thus, in addition to the pristine \ch{MoSe2/V2O5}, heterostructures with a defective TMD are also studied: \ch{(MoSe2}+\ch{V_{Se})/V2O5} and \ch{(MoSe2}+\ch{V_{Mo})/V2O5}. The synthesis of TMD is always accompanied by the occurrence of point defects in the structure, which can alter their properties~\cite{lin2016defect}. Defective layers have different thickness than the pristine ones, the defective sites may attract substitutions from other atomic species, and their adsorption abilities may also be different. It has been suggested that TMD monolayer may exhibit magnetic properties as a result of native defects as a result of the synthesis process ~\cite{lin2019defects, ma2011electronic, PhysRevB.95.245435, zhang2015strain}.  

First principles studies have shown that generally, in a 2D TMDs \ch{MX2} systems (M = W, Mo, Ni, Pt; X = Se, S, Te etc.) the chalcogen vancancy alone, which is the most stable defect type, cannot induce any local magnetic moment~\cite{thi2022enhanced, freire2022vacancy, khan2017electronic, yang2019electronic}. It may be possible, however, that transition metal vacancies may lead to a nonzero magnetic moment. For example, a \ch{MoSe2} monolayer with a Mo vacancy may exhibit a ferromagnetic state. This result is dependent on the distance between the defects, but also on the details of the simulations as demonstrated in recent studies ~\cite{lin2019defects, ma2011electronic, PhysRevB.95.245435, shafqat2017dft}. Based on our simulations, we find that the isolated \ch{MoSe2} without or with Se and Mo vacancies and the standalone
\ch{V2O5} do not display any magnetic properties. However, ferromagnetism is obtained in the  standalone defective substrate with a total magnetic moment of 2 $\mu_B$ after relaxation~\cite{xiao2009structural}.

The properties of \ch{V2O5} can also be altered by the introduction of oxygen vacancies V$_O$, resulting in formation of V--O--V bonds between the layers and making it ferromagnetic.  Oxygen vacancies can occur spontaneously during the synthesis process, or they can be created intentionally via other methods, including annealing or chemical reduction~\cite{scanlon2008ab, temsamani2020substrate, correal2023tuning, xiao2009structural}. Recently, thin films of \ch{V2O5} have been synthesized and theoretically studied~\cite{scanlon2008ab, das2019structural}. To understand how the defective substrate affects the system, we also consider $MoSe_2/(V_2O_5+V_O)$ in which the topmost \ch{V2O5} layer hosts an Oxygen vacancy as shown in Fig.~\ref{fig:structure}~(d).

For the systems studied here, we examine the total DFT energy for the FM state $E_{FM}$ in comparison with the total DFT energy for the state without magnetic ordering $E_{NM}$. We find that each heterostructure displays a preferred ferromagnetism, such that $E_{FM} - E_{NM}$~=~-0.46 meV/atom, -1.05 meV/atom, -1.492 meV/atom and -7.814 meV/atom for \ch{MoSe2/V2O5}, \ch{(MoSe2}+\ch{V_{Se})/V2O5}, \ch{(MoSe2}+\ch{V_{Mo})/V2O5} and \ch{MoSe2/(V2O5}+\ch{V_{O})}, respectively. Our computational results are in line with recent experiments~\cite{Kapuruge2023} which have reported that \ch{MoSe2/V2O5} systems indeed display room temperature ferromagnetism as further discussed below.

\begin{figure} [H]
    \centering
    \includegraphics[width=0.8\linewidth]{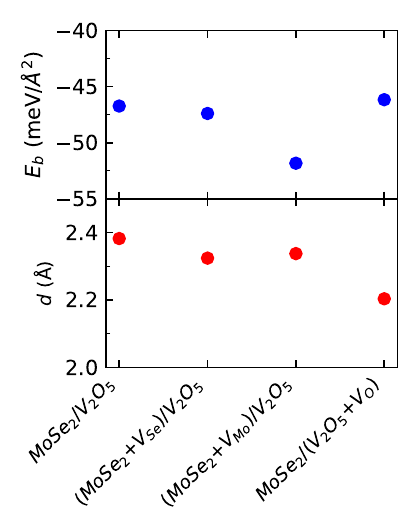}
    \caption{Binding energies $E_b$ and interlayer distances $d$ between the TMD monolayer and \ch{V2O5} substrate of the different heterostructures.}
    \label{fig:binding_energy}
    \vspace{1mm}
    \hrule
\end{figure}

The structural stability of each heterostructure with FM ordering is further examined by calculating the binding energies of the system, such that 
\begin{equation}
    E_{b} = \frac{E_{HST} - E_{TMD} - E_{Sub}}{A},
\end{equation}
where $E_{HST}$ is the total  energy of the heterostructure,  $E_{TMD}$ is the total  energy of isolated TMD monolayer, $E_{Sub}$ is the total  energy of the V$_2$O$_5$ substrate, and $A$ is the interfacial area in $\AA^2$ of the supercell. The results in Fig.~\ref{fig:binding_energy} show that all considered systems are stable with $E_b$ in the range of van der Waals attraction (10 to 100~meV/$\AA^2$)~\cite{PhysRevLett.110.263201, moo2019comprehensive}. 
We find that the most stable configuration is for the defective TMD monolayer with  Mo vacancy for which $E_b\sim-51$~meV/$\AA^2$ with 
 an interlayer separation d=2.33~$\AA$. 
The removal of Mo creates three dangling Se bonds at the interface that can have significant ionic hybridization with the substrate resulting in larger binding energy compared to the pristine heterostructure. For the case of
\ch{MoSe2/(V2O5}+\ch{V_{O})}, however, the binding energy $E_b=-46.16$~meV/$\AA^2$ is the weakest, but the interlayer distance $d=2.20$~$\AA$ is the shortest among the studied heterostructures. The removal of O from the topmost layer of the substrate breaks the V-O bond and creates a localized V cluster whose dangling bonds pull the TMD closer. However, the weaker hybridization due to the V cluster being buried inside the substrate and the short-ranged Pauli repulsion result in weaker $E_b$ compared to the other heterostructures.

Thus the results in Fig.~\ref{fig:binding_energy} show that stronger coupling does not necessarily correlate with smaller separation, as typically expected~\cite{PhysRevLett.110.263201, Zhang2013}.  Although $E_b$ is consistent with typical values of van der Waals energies~\cite{moo2019comprehensive}, the distance separation appears to be smaller than the typical van der Waals range of $3-4$~$\AA$. As discussed above, this  indicates that stronger chemical effects may be important for the stability of heterostructure.

\begin{figure} [H]
    \centering
    \includegraphics[width=0.7\linewidth]{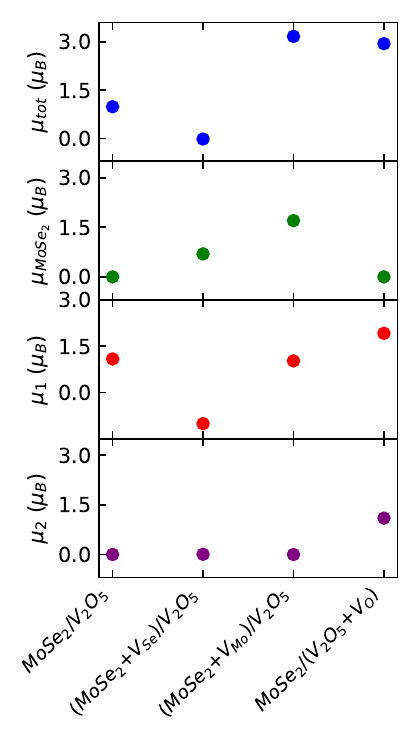}
    \caption{Magnetic moments for the studied heterostructures:  (a) the entire system, $\mu_{tot}$; (b) the \ch{MoSe2} monolayer, $\mu_{MoSe_2}$; (c) the topmost, $\mu_1$ and (d) middle layers, $\mu_2$ of the \ch{V2O5} substrate.}
    \label{fig:magnetic_mom}
    \vspace{1mm}
    \hrule
\end{figure}
We further discuss the simulations about the magnetization properties of the considered systems, shown in Fig.~\ref{fig:magnetic_mom}. We find that several heterostructures experience ferromagnetism including the pristine \ch{MoSe2/V2O5} for which $\mu_{tot}=1~\mu_B$ and it is carried almost exclusively by the topmost layer of the substrate $\mu_1=1.09$~$\mu_B$. In the heterostructure with Se vacancy, the total magnetization is almost zero ($\mu_{tot}=-0.01$~$\mu_B$). The \ch{MoSe2} layer carries magnetic moment of $\mu_{MoSe_2} = 0.7$~$\mu_B$, and the top layer of the substrate carries magnetic moment of $\mu_{1}=-1$~$\mu_B$. Fig.~\ref{fig:magnetic_mom} further shows that in the case of \ch{(MoSe2}+\ch{V_{Mo})/V2O5}, the total magnetization $\mu_{tot} = 3.17$~$\mu_B$ is higher than the one for the pristine structure and it is the largest. It appears that in this case both, the TMD and the topmost substrate layer, become magnetic with $\mu_{MoSe_2} = 1.7$~$\mu_B$ and $\mu_1 = 1.025$~$\mu_B$ while the second layer has a minimal contribution $\mu_2 = -0.004$~$\mu_B$. Finally, for the heterostructure with a defective substrate, the total magnetization is also significant ($\mu_{tot} = 2.95$~$\mu_B$) and it is determined exclusively by $\mu_1 = 1.91$~$\mu_B$ and $\mu_2 = 1.09$~$\mu_B$. 

Obtaining ferromagnetism in these systems is quite surprising.  Despite that \ch{MoSe2}, the defective TMD monolayer and the \ch{V2O5} display non-magnetic behavior, yet bringing the TMD and \ch{V2O5} substrate together leads to a nonzero net magnetization entirely contributed by the layers forming the heterostructure interface. Recent experimental results also confirm that robust room temperature ferromagnetism is observed in \ch{MoSe2/V2O5} heterostructures despite the nonmagnetic behavior of the TMD and the \ch{V2O5} substrate being individually nonmagnetic~\cite{Kapuruge2023}. Indeed, the reported M-H hysteresis behavior clearly indicates the formation of long-ranged magnetic ordering at room temperature and beyond. Our simulations not only show the emergent interface magnetism in \ch{MoSe2/V2O5} heterostructures, but they also reveal the underlying reasons for this phenomenon.

\subsection*{Electronic structure properties}
\begin{figure}
    \centering
    \includegraphics[width=0.85\linewidth]{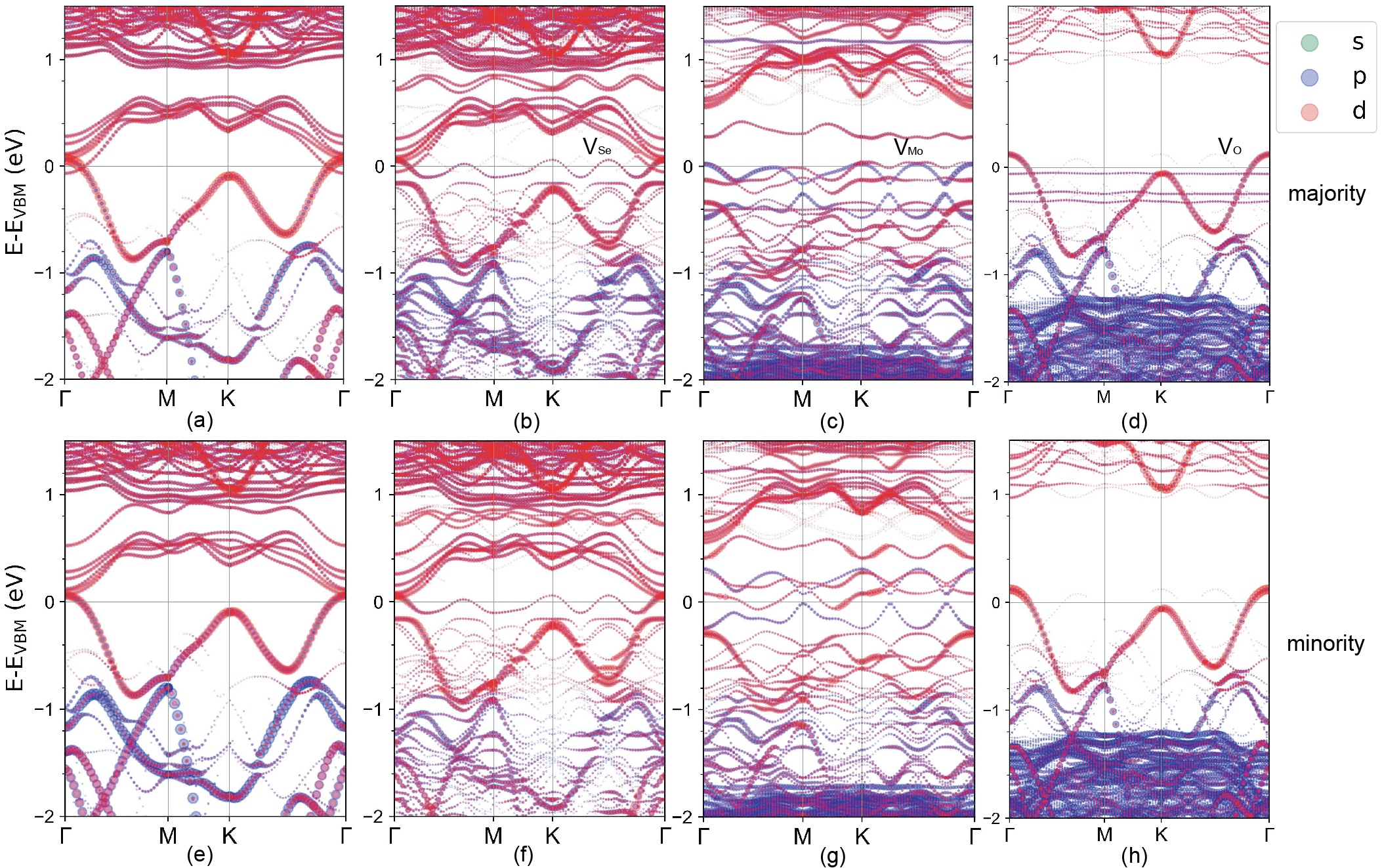}
    \caption{Orbitally resolved electronic band structures for: (a) majority and (e) minority carriers of \ch{MoSe2/V2O5}; (b) majority and (f) minority carriers of \ch{(MoSe2}+\ch{V_{Se})/V2O5}; (c) majority and (g) minority carriers of \ch{(MoSe2}+\ch{V_{Mo})/V2O5}; (d) majority and (h) minority carriers of \ch{MoSe2/(V2O5}+\ch{V_{O})}. }
    \label{fig:enter-electronic}
    \vspace{1mm}
    \hrule
\end{figure}

The origin of the interface magnetism in the \ch{MoSe2/V2O5} heterostructures is further analyzed in terms of the computed electronic band structure. In Fig.~\ref{fig:enter-electronic}, results are shown for the energy bands projected on the Brillouin zone of the TMD layer for the majority and minority carriers delineating between the $s$, $p$, and $d$ orbitals. 

Despite the nonmagnetic nature of the TMD monolayer and substrate, the interface in the \ch{MoSe2/V2O5} heterostructure leads to energy band reconstruction: near the Fermi level, there are bands responsible for the minority valence band maximum that are composed of Se-$p$ and Mo-$d$ hybridized orbitals with major contribution coming from Mo-$d_{z^2}$. The conduction band minimum is due to V-$d$ and O-$p$ states, with major contribution coming from V-$d_{xy}$ orbital. (The band structures projected on the orbitals of the composing elements are provided in the Fig.~S3-S4). The energy band composition for the majority spin carriers is similar, although there are notable distinctions. It is found that the degeneracy of the majority conduction bands at the $\Gamma$ point is lifted which results in a larger energy window spread of 0.4 eV compared to the minority carriers for which this range is 0.3 eV. Differences of the majority and minority bands are also found at the $K$-point with more continuous energy spread seen in the minority conduction band than the majority conduction band. A further  distinction is the appearance a minority energy band crossing $\Gamma$ at 0.5 eV composed of strong hybridization between V-$d$ and O-$p$.

Fig.~\ref{fig:enter-electronic}~(b, f) shows that the \ch{(MoSe2}+\ch{V_{Se})/V2O5} heterostructure also has metallic nature due to the appearance of a relatively dispersionless band at the Fermi level in both types of carriers. While Se-$p$ and Mo-$d_{z^2}, d_{x^2-y^2}$ orbitals mostly make up the valence bands near the Fermi level and the flat bands at the Fermi level, the lowest lying conduction bands are dominated by V-$d_{xy}$ and O-$p$ states (Fig.~S5 and S6). The majority and minority band structures differ in the $0.75-0.8$ eV region: although, for both carriers hybridized V-$d_{xy}, d_{xz}$ and O-$p$ orbitals are found, there are  additional bands for the minority carriers. There are also differences around the characteristic $\Gamma, K$ and $M$ points in the minority vs majority bands for energies less than 1 eV. 
\end{multicols}
\begin{multicols}{2}
Distinct minority and majority band structures are also found for the \ch{(MoSe2}+\ch{V_{Mo})/V2O5} heterostructure, as shown in Fig.  \ref{fig:enter-electronic}~(c, g). The flat bands near the Fermi level are defect-like states similar to the defect-like flat bands due to the Se vacancy in the \ch{(MoSe2}+\ch{V_{Se})/V2O5} system. The valance bands for both spin carriers are $p-d$ hybridized, composed mostly of Mo-$d_{x^2-y^2}, d_{z^2}$ and Se-$p$ states near the Fermi level and O-$p$ states dominating the core regions. The band crossing the Fermi level for both carriers are made up of strongly hybridized V-$d$ and O-$p$ orbitals with major contribution of V-$d_{xy}$ in the energy range 0.5--1.1~eV, and V-$d_{x^2-y^2}, d_{xz}$  with some minor contributions also coming from Mo-$d_{x^2-y^2}, d_{xy}, d_{z^2}$ and Se-$p$. The conduction bands for both the carriers in the  0.5--1.1~eV range contain hybridized V-$d_{xy},d_{x^2-y^2}, d_{xz}$ and O-$p$ states with some minor Mo-$d_{x^2-y^2}, d_{xy}, d_{z^2}$ and Se-$p$ admixture (Fig.~S7 and S8). The majority and minority band structures differ especially in the (-0.5, 0.5) eV energy range. For the majority carriers, relatively flat bands are found at $E_F$ and at $E\sim-0.4$~eV, while for the minority carriers conduction bands are found at and near $E_F$. It appears that the flat majority conduction band around $0.4$~eV corresponds to a similar band for the minority carriers by shifted upwards to around $0.5$~eV. 

The results for the \ch{MoSe2/(V2O5}+\ch{V_{O})} heterostructure in Fig.~\ref{fig:enter-electronic}~(d, h) show pronounced band signatures of the \ch{MoSe2} monolayer with a characteristic band gap at the K-point of 1.12~eV for both types of carriers which compares well to the energy gap of 1.03~eV for the monolayer (see Fig.~S1 for the monolayer band structure). A clear distinction between majority and minority carriers are the defect-like states in the valence region found only for majority spin polarization. These flat valence band bands near the Fermi level come from the strong hybridization of V-$d$ and O-$p$ orbitals of the majority carriers, as shown in Fig.~\ref{fig:enter-electronic}~(d, h). The flat band closest to the Fermi level contains V-$d_{xy}$ orbitals, the band at -0.3~eV has V-$d_{xz}$ orbitals and the band at -0.35~eV has V-$d_{yz}$ orbitals. For valence states in the $<-1$~eV energy region, strong Mo-$d$-Se-$p$ hybridization is found for both types of carriers. The conduction region $>1$~eV is composed of V-$d$, Mo-$d$, O-$p$, and Se-$p$ state admixture, again for both types of carriers (Fig.~S9 and S10).

Spin-orbit coupling (SOC) may also be important for the electronic structure of TMDs. Here we also include this relativistic correction in the calculations of the energy band structure of the heterostructures and the results are shown in Fig.~S11. The SOC does not preserve the spin degree of freedom and the distinction between majority and minority carriers is not possible. Comparing with the results with no SOC, we find that for the pristine heterostructure, the SOC is responsible for shifting of several conduction bands upwards resulting in a band gap of 0.7~eV at $\Gamma$ and 0.9~eV at the K-point (Fig.~S11~(e)). The SOC is also responsible for lifting the degeneracy of the valence and conduction bands associated with the Mo atoms (especially prominent at the K-point where the valence band splitting is around 190 meV).
\end{multicols}
\begin{figure}
    \centering
    \includegraphics[width=0.87\linewidth]{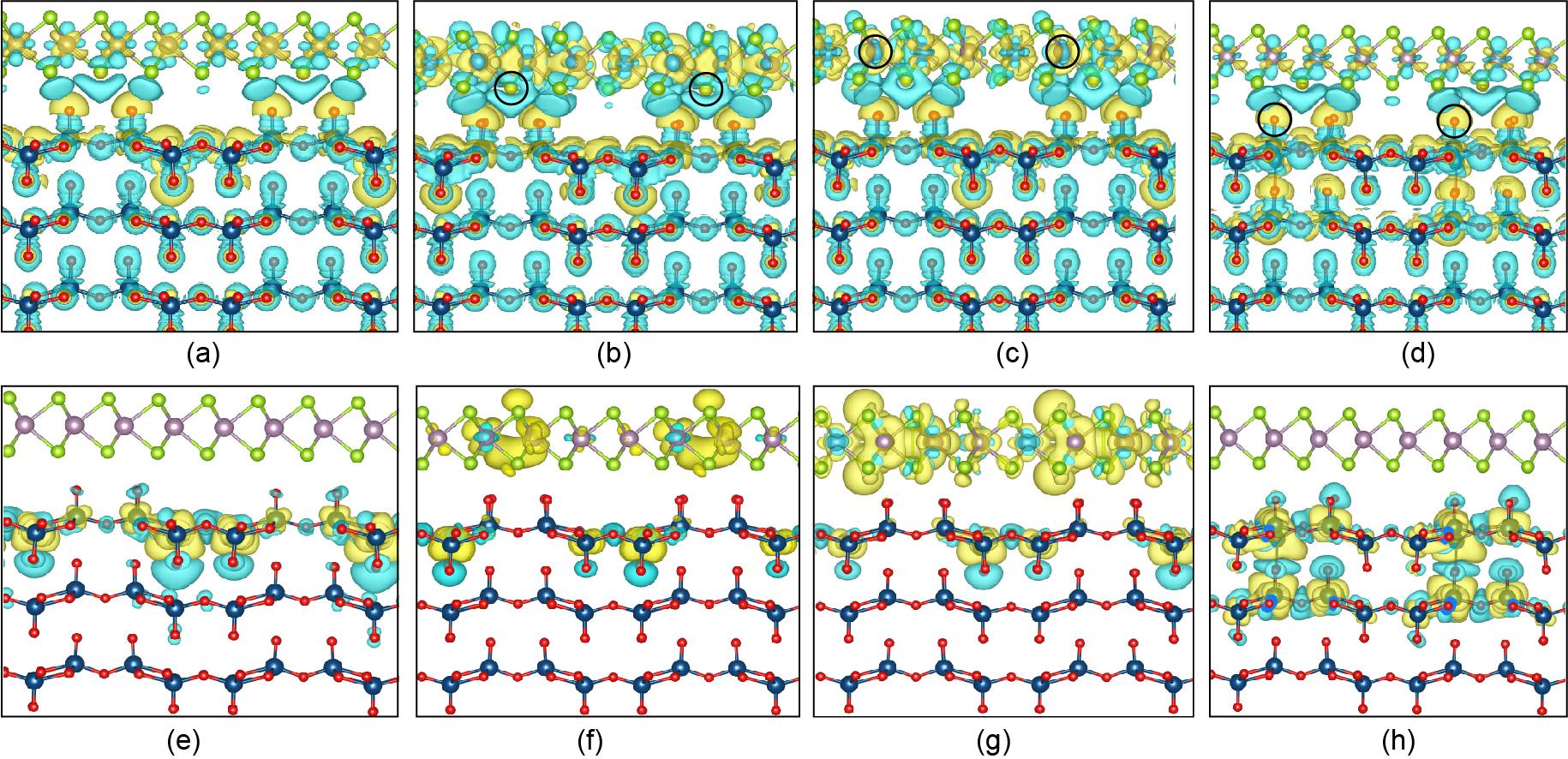}
    \caption{Charge Density Difference (blue-negative, yellow positive) of (a) \ch{MoSe2/V2O5} heterostructure. (b) \ch{(MoSe2}+\ch{V_{Se})/V2O5} heterostructure. (c) \ch{(MoSe2}+\ch{V_{Mo})/V2O5} heterostructure. (d) \ch{MoSe2} and \ch{(V2O5}+\ch{V_O)} heterostructure. The isosurface value is $7\times10^{-4}$~$e/\AA^3$ for all cases. Spin density (blue-negative, yellow positive) of (e) \ch{MoSe2/V2O5} heterostructure (isosurface value of $7\times10^{-5}$~$e/\AA^3$). (f) \ch{(MoSe2}+\ch{V_{Se})/V2O5} heterostructure (isosurface value of $7\times10^{-4}$~$e/\AA^3$). (g) \ch{(MoSe2}+\ch{V_{Mo})/V2O5} heterostructure (isosurface value of $4\times10^{-4}$~$e/\AA^3$). (h) \ch{MoSe2/(V2O5}+\ch{V_{O})} heterostructure (isosurface value of $1\times10^{-4}$~$e/\AA^3$).}
    \label{fig:charge_density}
    \vspace{1mm}
    \hrule
\end{figure}
\begin{multicols}{2}
The valance bands are spin split everywhere in the Brillouin zone except at the high symmetry points $\Gamma$ and M where the bands remain degenerate, which is a typical SOC effect seen in the bands of monolayer TMDs~\cite{kormanyos2014spin, sun2016indirect, roldan2014momentum}. An appearance of a flat band at -0.18~eV due to Mo-$d$ orbital is also present.
Something similar is obtained for the \ch{(MoSe2}+\ch{V_{Se})/V2O5} system, however impurity bands due to the vacancy are also found at around 0.1~eV region in the conduction band and at -0.38~eV in the valance band (Fig.~S11~(f)) with valence band splitting of 190~meV at the K-point. The V-$d$ and O-$p$ hybridized conduction bands are shifted upward while the Mo-$d$ and Se-$p$ hybridized flat bands are shifted downward to 0.4--0.6~eV range. The band gap at $\Gamma$ point is 0.2~eV. In the case of the \ch{(MoSe2}+\ch{V_{Mo})/V2O5} and \ch{MoSe2/(V2O5}+\ch{V_{O})} heterostructures, however, the SOC does not result in significant band shifts (Fig.~S11~(g, h)). The band splitting is not present in the case of \ch{(MoSe2}+\ch{V_{Mo})/V2O5} heterostructure. The band structure of \ch{MoSe2/(V2O5}+\ch{V_{O})} heterostructure is very similar to the band structure of pristine heterostructure with valence band splitting of 180~meV at K-point and the bands remaining degenerate at $\Gamma$ and M.

\subsection*{Charge density difference and spin redistribution}
We further analyze the charge density difference and spin density distribution for each heterostructure with results shown in Fig.~\ref{fig:charge_density}. The charge density difference is calculated by $\Delta \rho = \rho_{HST} - (\rho_{TMD} + \rho_{substr})$, where $\rho_{HST}$, $\rho_{TMD}$ and $\rho_{substr}$ are the total charges in the heterostructure, isolated TMD monolayer and isolated \ch{V2O5} substrate, respectively. Charge is transferred from the negative regions of $\Delta \rho$ signifying depleted regions (blue isosurfaces) to the positive regions signifying regions with accumulated electrons (yellow isosurfaces).

The charge density difference in the \ch{MoSe2/V2O5} heterostructure in Fig.~\ref{fig:charge_density}~(a) shows that depleted electrons from the substrate atoms and the lower plane of Se atoms of the TMD  monolayer are accumulated on the top layer and the terminally bound O atom on the top layer. It appears that there is a layer of electronic charge distribution on the top layer of \ch{V2O5} substrate and a layer of positive charge distribution on the Se atoms. As a result, there is an increased electron density at the top layer of substrate which is responsible for the metallic nature of the heterostructure. This charge transfer between the neutral substrate and TMD monolayer is made possible due to the difference in their Fermi energy levels as the charge gets transferred from the higher to the lower Fermi level to reach equilibrium. The increased electron density on the substrate results in unpaired V-$d$ electrons which is the source of magnetic moments. 

Tracking the spin degree of freedom in Fig.~\ref{fig:charge_density}~(e) shows that the O atoms in the gap carry the majority spins, while the minority spins are carried by the Vanadium atoms in the topmost layer of the substrate. The net magnetization of $\mu_{tot}$~=~1~$\mu_B$ is mostly contributed by the minority carriers (blue isosurfaces) in the top layer of the substrate while a small contribution of majority carriers are also present on the second layer. 

For the defective \ch{(MoSe2}+\ch{V_{Se})/V2O5} system, we find that its charge redistribution in Fig.~\ref{fig:charge_density}~(b) is different than the pristine heterostructure, particularly on the TMD layer as the charge is being redistributed on \ch{MoSe2} with charge depleting from the Se atoms to the Mo atoms. Significant charge depletion from the lower plane of Se atoms of \ch{MoSe2} monolayer to the substrate is also seen. The regions of accumulated and depleted charges are concentrated mostly around the vacancy location. The spin redistribution in the defective heterostructure shows that the TMD is polarized, consisting mainly of majority spin carriers contributing to a magnetic moment of $\mu_{MoSe2}=0.7$~$\mu_B$. (Fig.~\ref{fig:charge_density}~(f)). On the substrate, the spin polarization is prominently seen in the top layer with a net magnetization of $\mu_{tot}=-0.01$~$\mu_B$ associated with the minority carriers (blue isosurfaces) in top substrate layer and the TMD layer.

The charge redistribution of the \ch{(MoSe2}+\ch{V_{Mo})/V2O5} heterostructure is similar to the Se-defective case (Fig.~\ref{fig:charge_density}~(c)), since electrons are accumulated at the top substrate layer with depletion regions concentrated just below the vacancy sites in the TMD monolayer. The spin density, however, shows notable differences. As given in Fig.~\ref{fig:charge_density}~(g), we  find that the TMD monolayer bares most of the majority spin distribution contributing to a net magnetization of $\mu_{MoSe_2} = $~1.7~$\mu_B$. The overall net magnetization of $\mu_{tot} = $~2.724~$\mu_B$ comes collectively from the TMD and the top layer of the substrate.

The heterostructure with the defective substrate shows charge redistribution in Fig.~\ref{fig:charge_density}~(d) that is similar to that of the pristine heterostructure. The charge depletion occurs from the substrate and the lower plane of Se atoms of the TMD monolayer and accumulates on the top layer and the terminally bound O atom on the top and middle layer of the substrate. The spin redistribution in this case shows significant distinction. The spin polarization unlike the other cases is not only on the top layer but is almost equally dominant in the first and second layer with the majority spin carries (yellow isosurfaces) contributing to the total magnetic moment of $\mu_{tot}=2.95$~$\mu_B$. The  majority and minority spin accumulation is located around the location of the O vacancy. 

It is not uncommon for 2D TMDs to bear ferromagnetic ordering due to proximity to magnetic substrate as a van der Waals heterostructures~\cite{zhang2016large}, however in this work we have achieved magnetic ordering in a system where constituting materials are non magnetic. Our results show that bringing these materials together can facilitate charge transfer which permits spin redistribution among the van der Waals materials giving rise to ferromagnetic ordering. Such phenomenon is made possible in \ch{MoSe2} and \ch{V2O5} due to favorable type-III band alignment of these two materials making \ch{MoSe2} p-doped. As shown in a previous study of this phenomenon~\cite{arora2025engineering}, such band alignment allows spontaneous charge transfer between 2D materials and binary oxide substrates. With this alignment allowing charge transfer through the interface along with an unequal distribution of the two types of carriers, we can expect the possibility of similar phenomenon of interface magnetism in families of TMDs and binary oxides that exhibits type-III band alignment.

\section*{Conclusion}
Ferromagnetic nanostructured systems are of great interest for fundamental discoveries and new application development. Here we have shown the existence of  ferromagnetism in \ch{MoSe2/V2O5} systems. Despite that each component, the \ch{MoSe2} monolayer and the \ch{V2O5} thin film are nonmagnetic, their heterostructure is ferromagnetic. Native point defects, unavoidable in the synthesis process, can enhance or inhibit the magnetism as we have shown in the case of the most commonly found Se, Mo, and O point defects. The emergent interface magnetism arises from hybridization between various atomic orbitals accompanied by charge transfer and spin redistribution in the interface region. Our computational studies give the theoretical framework of the recently observed ferromagnetic \ch{MoSe2/V2O5} systems at room temperature~\cite{Kapuruge2023}.

The heterostructure studied here is another example similar to the case of the ferromagnetic \ch{LaAlO3/SrTiO3} thin films, where the individual oxides are nonmagnetic~\cite{song2021electronic, hu2016oxygen, mohanta2014oxygen}. Changing the carrier concentration and O defect formation can modify the magnetic response of  \ch{LaAlO3/SrTiO3}. In the case of \ch{MoSe2/V2O5}, different defects  can also be used to control the interface magnetism. It is also interesting to consider how the concentration of defects can affect the magnetic properties of these systems.
\ch{MoSe2/V2O5} heterostructures broaden the materials library of systems with sustainable magnetism
confined at atomically sharp interfaces, which can be
useful for spintronic device applications at the nanoscale.

\section*{Acknowledgments}
We acknowledge support from the US Department of Energy under Grant No. DE-FG02-06ER46297. Computational resources were provided by USF Research Computing. We acknowledge discussions with Profs. Humberto Rodriguez Gutierrez (University of South Florida) and Douglas Galvao (State University of Campinas).

\section*{Author information}
\subsection*{Corresponding Author}
\begin{itemize}
    \item Lilia M. Woods - Department of Physics, University of South Florida, Tampa, Florida 33620, United States; \href{https://orcid.org/0000-0002-9872-1847}{orcid.org/0000-0002-9872-1847}; Email: \href{mailto:lmwoods@usf.edu}{lmwoods@usf.edu}.
\end{itemize}
 
\subsection*{Authors}
\begin{itemize}
    \item Rohin Sharma - Department of Physics, University of South Florida, Tampa, Florida 33620, United States; \href{https://orcid.org/0000-0003-4387-1642}{orcid.org/0000-0003-4387-1642}.
    \item Diem Thi-Xuan Dang - Department of Physics, University of South Florida, Tampa, Florida 33620, United States; \href{https://orcid.org/0000-0001-7136-4125}{orcid.org/0000-0001-7136-4125}.
\end{itemize}
\vspace{3mm}
\hrule
\vspace{3mm}

\bibliographystyle{ieeetr}
\end{multicols}
\end{document}


\maketitle

\begin{figure} [H]
    \centering
    \includegraphics[width=0.8\linewidth]{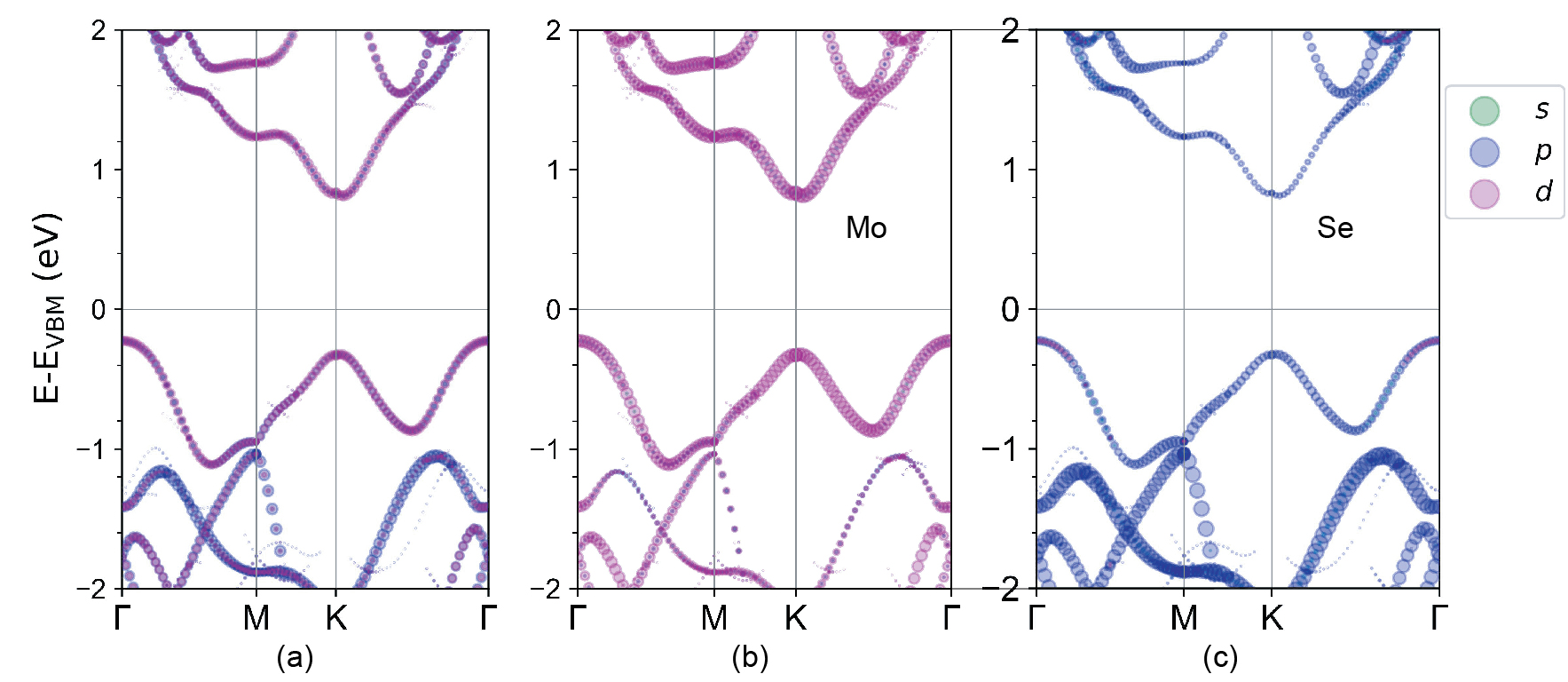}
    \caption{(a) Band structure of $MoSe_2$ monolayer; (b) Contributions of Mo orbitals; (c) Contributions of Se orbitals.}
    \label{fig:SI_MoSe2 band}
    \vspace{1mm}
    \hrule 
\end{figure}

The orbitally projected energy band structure of the $MoSe_2$ monolayer is shown in Fig.~\ref{fig:SI_MoSe2 band}. We find that the TMD is a direct semiconductor with an energy gap of 1.03~eV at the K-point of the Brillouin zone.

\begin{figure} [H]
    \centering
    \includegraphics[width=0.8\linewidth]{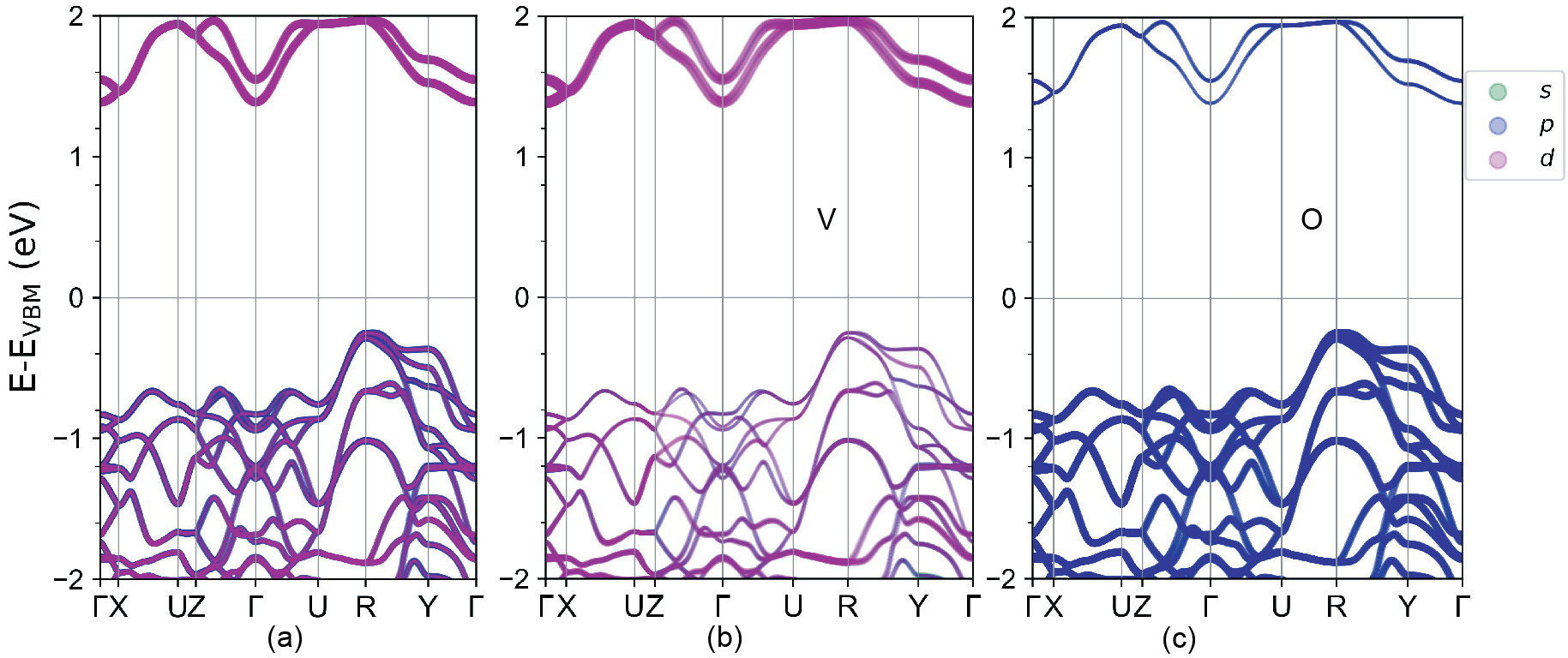}
    \caption{(a) Orbitally projected band structure of the $V_2O_5$ substrate; (b) Contributions of orbitals associated with the V atom; (c) Contributions of orbitals associated with the O atom.}
    \label{fig:SI_bulk band}
    \vspace{1mm}
    \hrule 
\end{figure}

The orbitally projected band structure for the $V_2O_5$ substrate as considered in the main text is shown in Fig.~\ref{fig:SI_bulk band}. We find that the substrate is an indirect semiconductor with an energy gap of 1.64~eV.

The orbitally projected band structure projected on the TMD Brillouin zone for the pristine $MoSe_2/V_2O_5$ heterostructure with resolved majority and minority spin contributions. Signatures from the isolated TMD monolayer in Fig.~\ref{fig:SI_MoSe2 band} show that $MoSe_2$ remains a semiconductor with a gap of 1.151~eV at the K-point. 
\begin{figure} [H]
    \centering
    \includegraphics[width=0.8\linewidth]{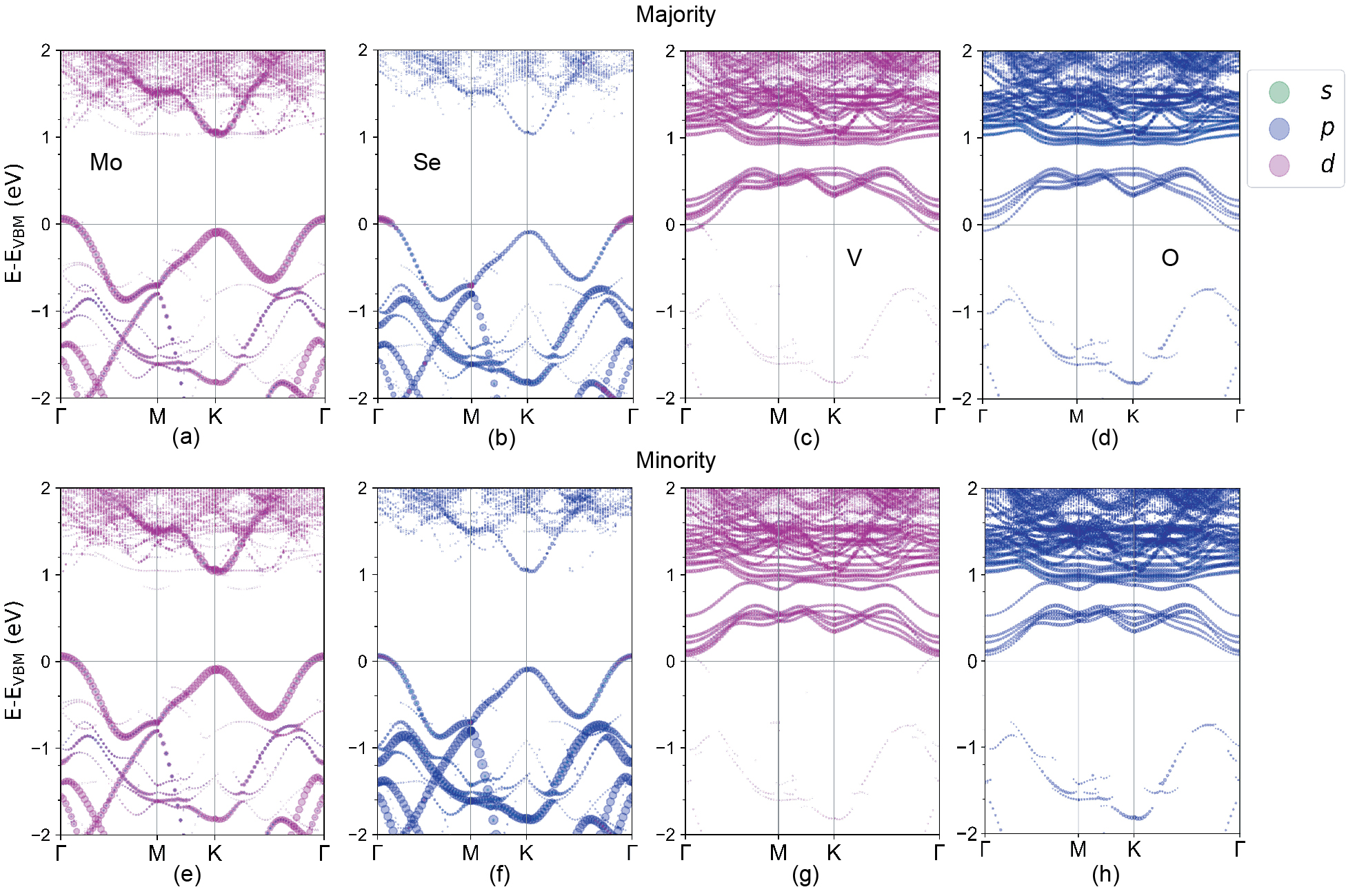}
    \caption{Orbitally projected band structure of the pristine $MoSe_2/V_2O_5$ heterostructure in the Brillouin zone of the TMD monolayer for (a) majority and (b) minority carriers. Contributions from Mo atoms for (c) majority and (d) minority carriers. Contributions from V atoms for (c) majority and (d) minority carriers. Contributions from O atoms for (c) majority and (d) minority carriers.}
    \label{fig:SI_No vac band}
    \vspace{1mm}
    \hrule 
\end{figure}
\begin{figure} [H]
    \centering
    \includegraphics[width=0.65\linewidth]{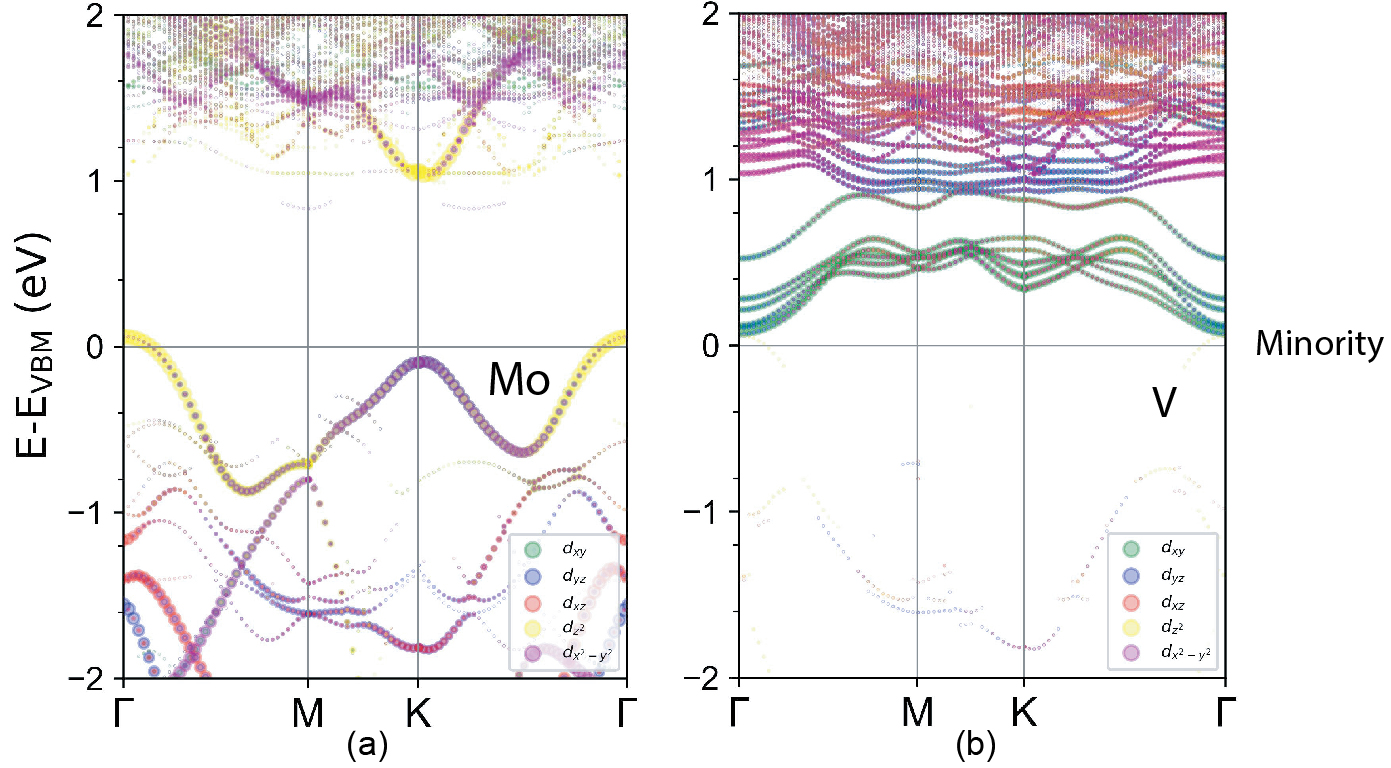}
    \caption{Breakdown of the $d$ orbital contribution of $Mo$ and $V$ of the heterostructure in the minority charge carriers.}
    \label{fig:d_orb_band_novac}
    \vspace{1mm}
    \hrule 
\end{figure}

The orbitally projected band structure projected on the TMD Brillouin zone for the $MoSe_2+V_{Se}/V_2O_5$ heterostructure with resolved majority and minority spin contributions. Several defect-like state at and near the Fermi level show that the semiconducting nature os $MoSe_2$ is not preserved.

\begin{figure} [H]
    \centering
    \includegraphics[width=0.8\linewidth]{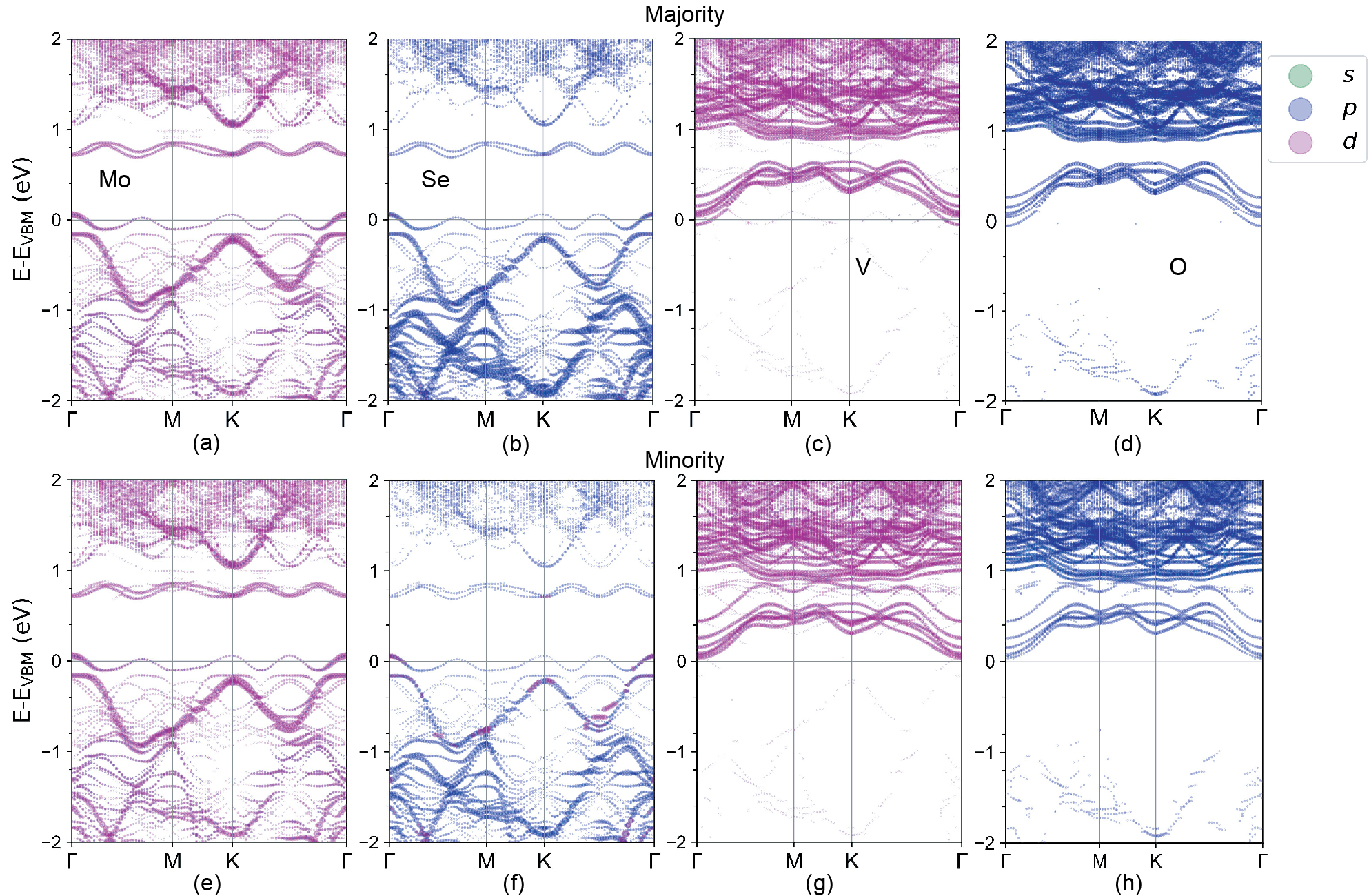}
    \caption{Orbitally projected band structure of the $MoSe_2+V_{Se}/V_2O_5$ heterostructure in the Brillouin zone of the TMD monolayer for (a) majority and (b) minority carriers. Contributions from Mo atoms for (c) majority and (d) minority carriers. Contributions from V atoms for (c) majority and (d) minority carriers. Contributions from O atoms for (c) majority and (d) minority carriers.}
    \label{fig:SI_Se vac band}
    \vspace{1mm}
    \hrule 
\end{figure}
\begin{figure} [H]
    \centering
    \includegraphics[width=0.63\linewidth]{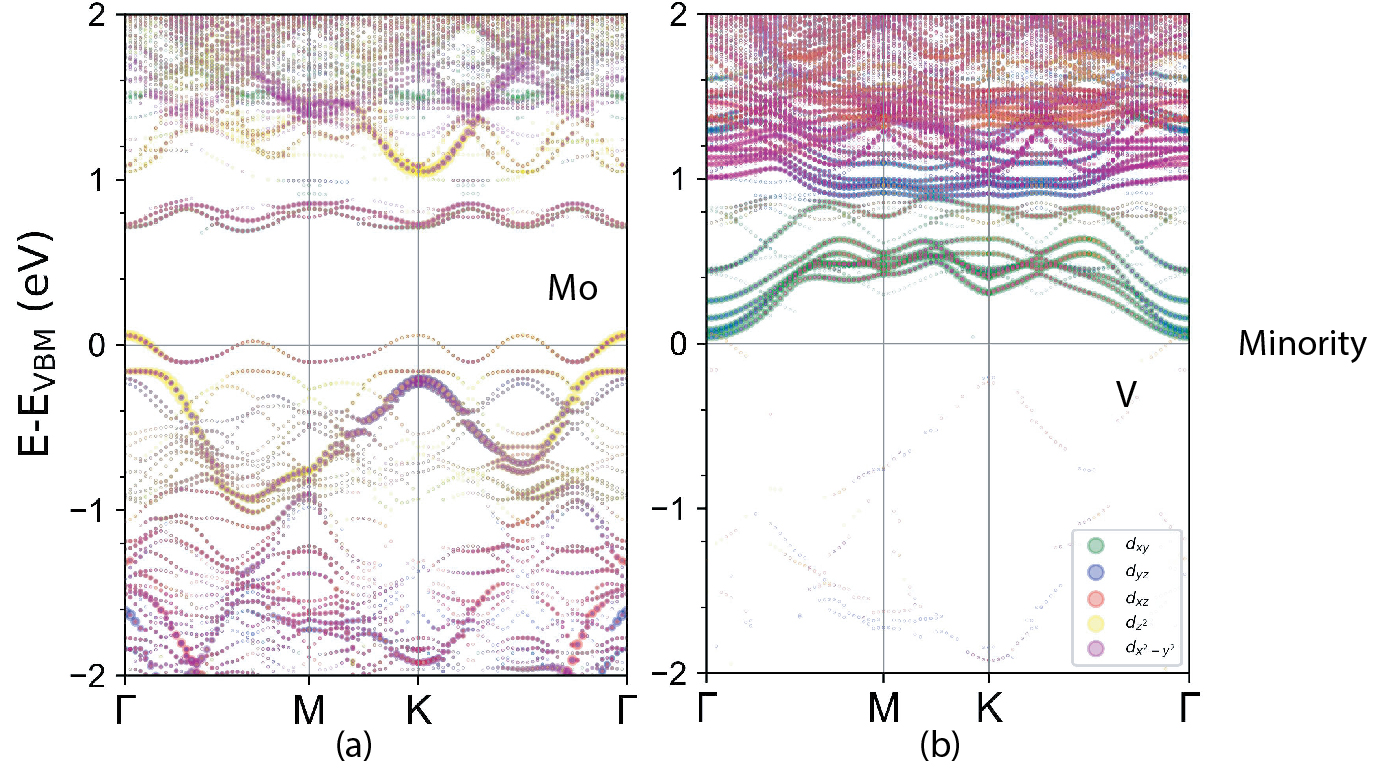}
    \caption{Breakdown of the $d$ orbital contribution of $Mo$ and $V$ of the heterostructure with Se vacancy in the minority charge carriers.}
    \label{fig:d_orb_band_sevac}
    \vspace{1mm}
    \hrule 
\end{figure}
The orbitally projected band structure projected on the TMD Brillouin zone for the 
 $MoSe_2+V_{Mo}/V_2O_5$ heterostructure with resolved majority and minority spin contributions. Several defect-like state at and near the Fermi level show that the semiconducting nature os $MoSe_2$ is not preserved.

\begin{figure} [H]
    \centering
    \includegraphics[width=0.8\linewidth]{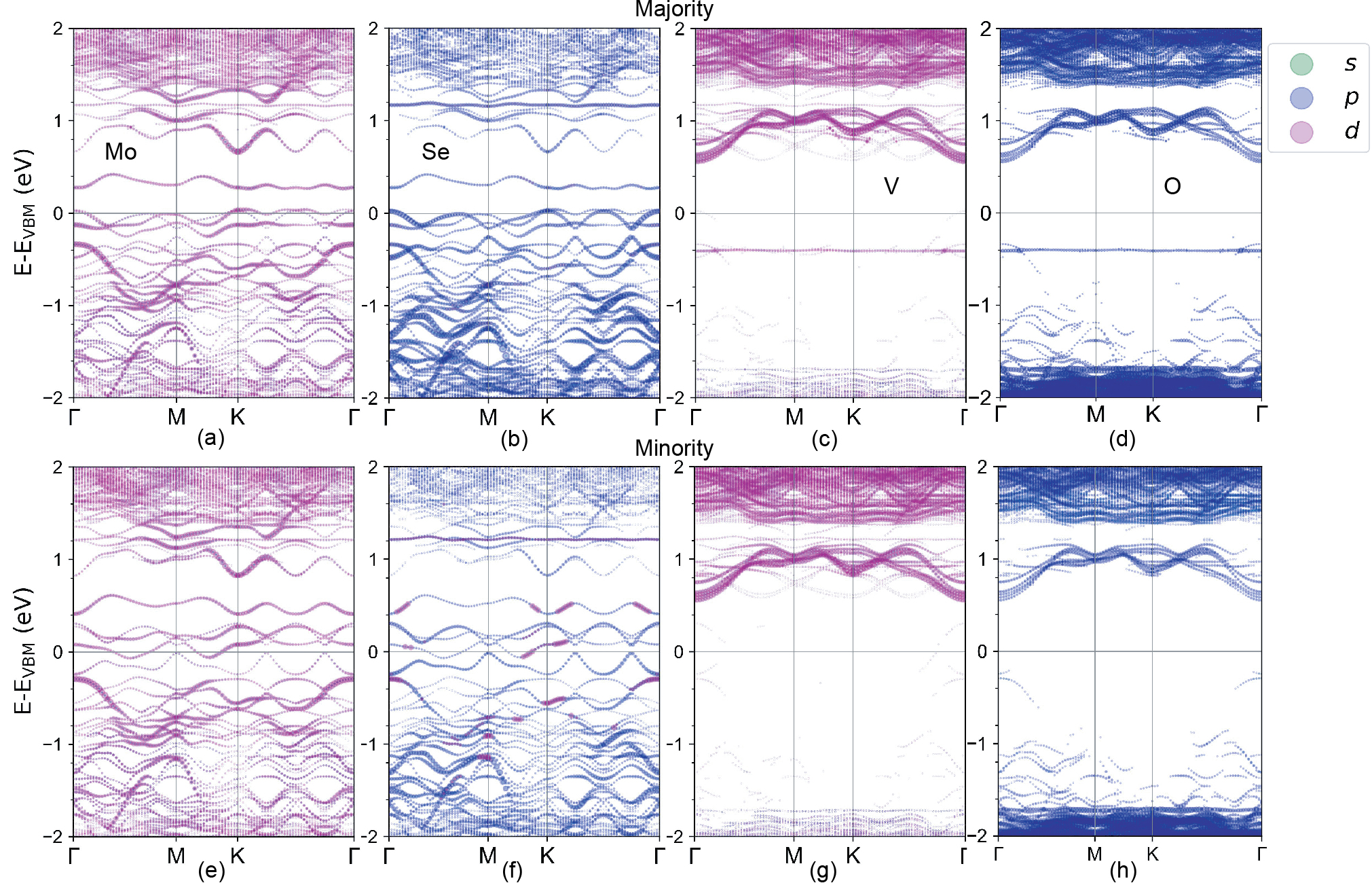}
    \caption{Orbitally projected band structure of the $MoSe_2+V_{Mo}/V_2O_5$ heterostructure in the Brillouin zone of the TMD monolayer for (a) majority and (b) minority carriers. Contributions from Mo atoms for (c) majority and (d) minority carriers. Contributions from V atoms for (c) majority and (d) minority carriers. Contributions from O atoms for (c) majority and (d) minority carriers.}
    \label{fig:SI_Mo vac band}
    \vspace{1mm}
    \hrule 
\end{figure}
\begin{figure} [H]
    \centering
    \includegraphics[width=0.63\linewidth]{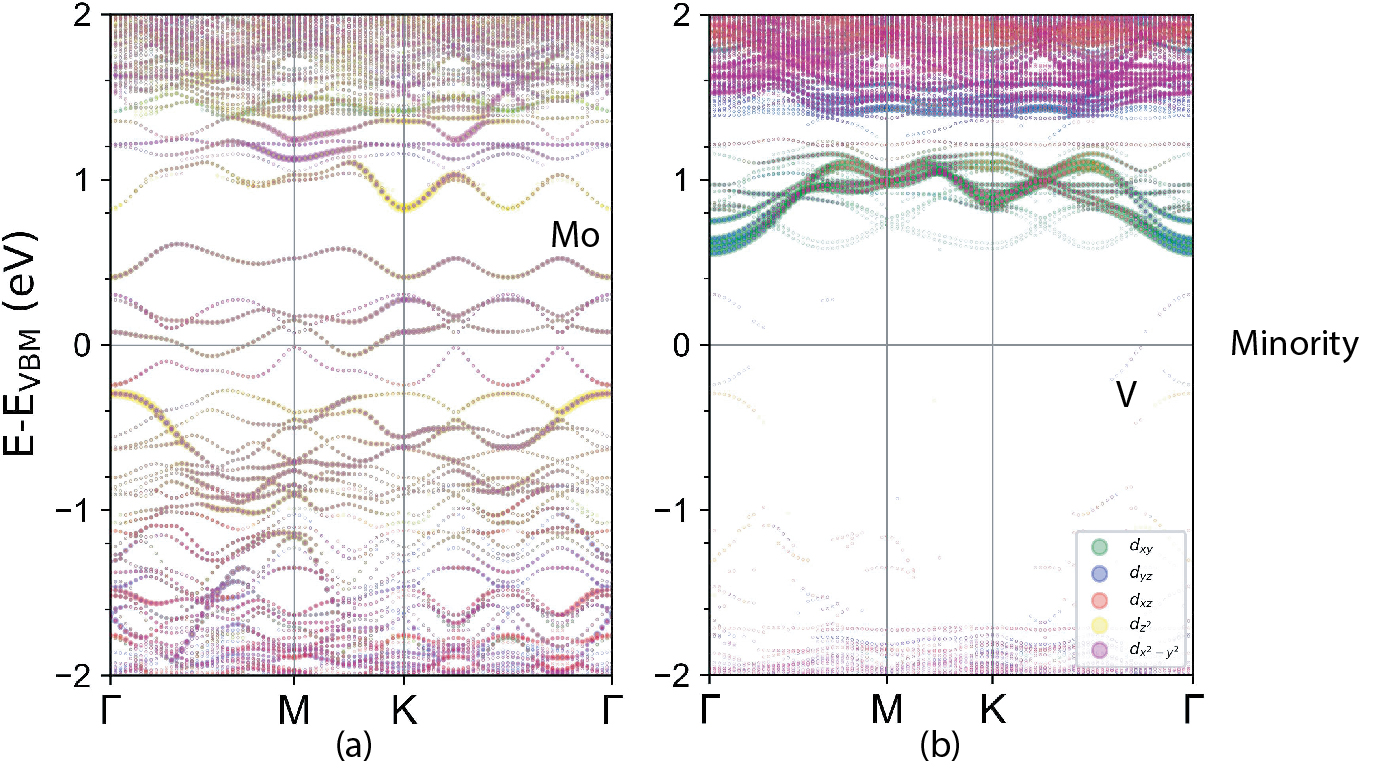}
    \caption{Breakdown of the $d$ orbital contribution of $Mo$ and $V$ of the heterostructure with Mo vacancy in the minority charge carriers.}
    \label{fig:d_orb_band_movac}
    \vspace{1mm}
    \hrule 
\end{figure}

The orbitally projected band structure projected on the TMD Brillouin zone for the $MoSe_2/V_2O_5+V_O$ heterostructure with resolved majority and minority spin contributions. The results indicate that $MoSe_2$ is a semiconductor with a gap of 1.12~eV at the K-point in the its Brillouin zone, however there are are defect-like states near $E_F$ associated with the substrate due to the O vacancy.
\begin{figure} [H]
    \centering
    \includegraphics[width=0.8\linewidth]{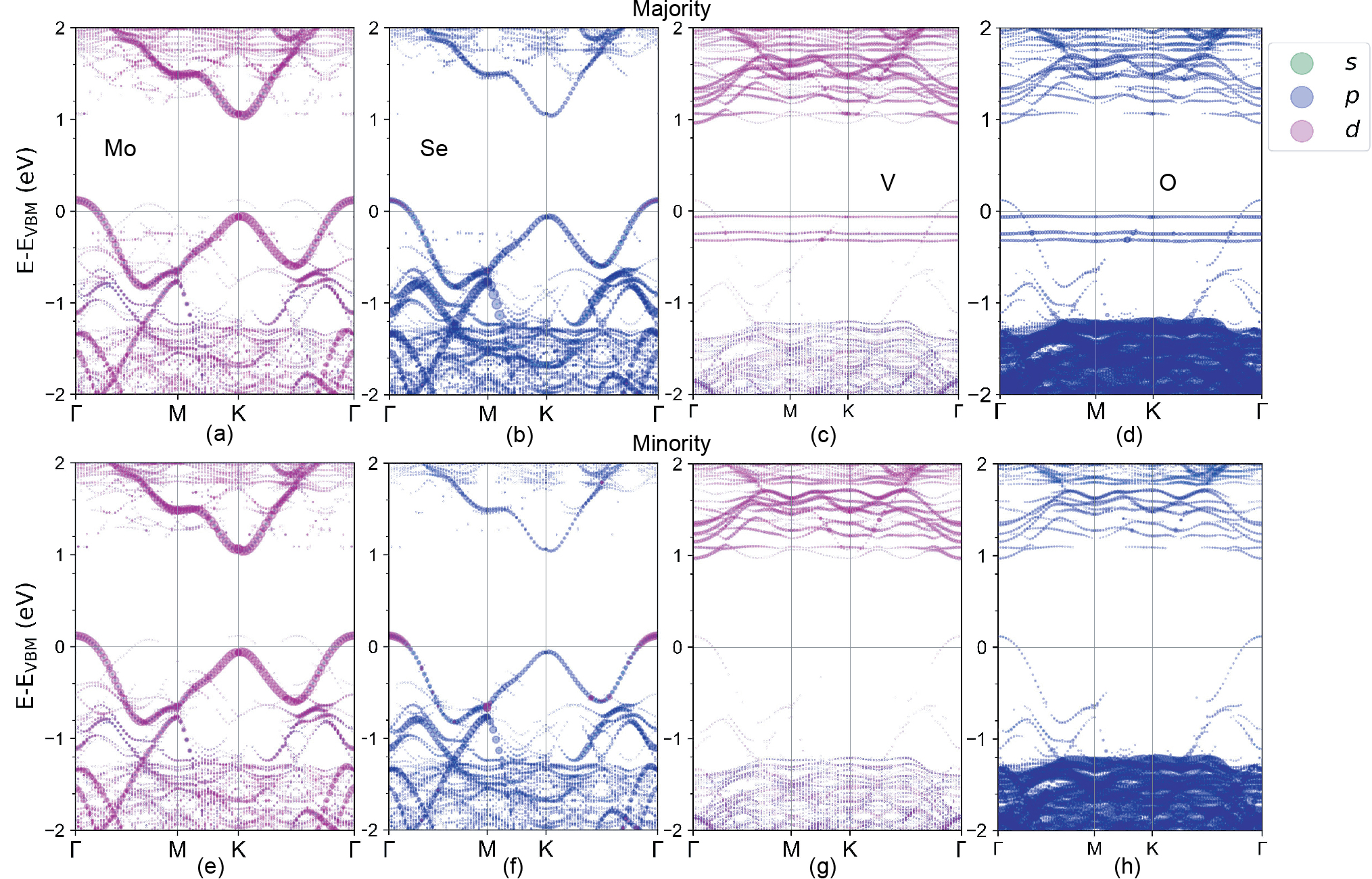}
    \caption{Orbitally projected band structure of the $MoSe_2/V_2O_5+V_O$ heterostructure in the Brillouin zone of the TMD monolayer for (a) majority and (b) minority carriers. Contributions from Mo atoms for (c) majority and (d) minority carriers. Contributions from V atoms for (c) majority and (d) minority carriers. Contributions from O atoms for (c) majority and (d) minority carriers.}
    \label{fig:SI_O vac band}
    \vspace{1mm}
    \hrule 
\end{figure}
\begin{figure} [H]
    \centering
    \includegraphics[width=0.63\linewidth]{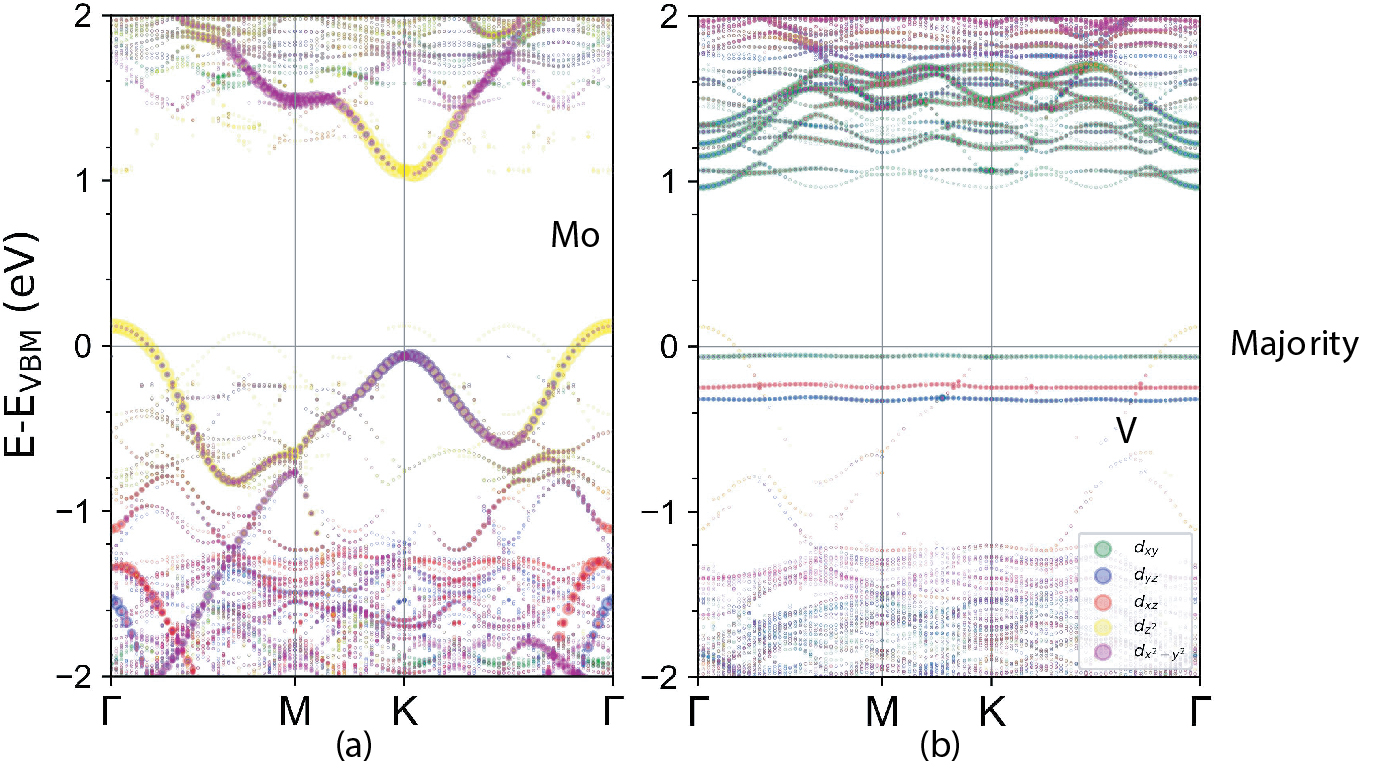}
    \caption{Breakdown of the $d$ orbital contribution of $Mo$ and $V$ of the heterostructure with O vacancy in the majority charge carriers.}
    \label{fig:d_orb_band_ovac}
    \vspace{1mm}
    \hrule 
\end{figure}

\begin{figure} [H]
    \centering
    \includegraphics[width=0.8\linewidth]{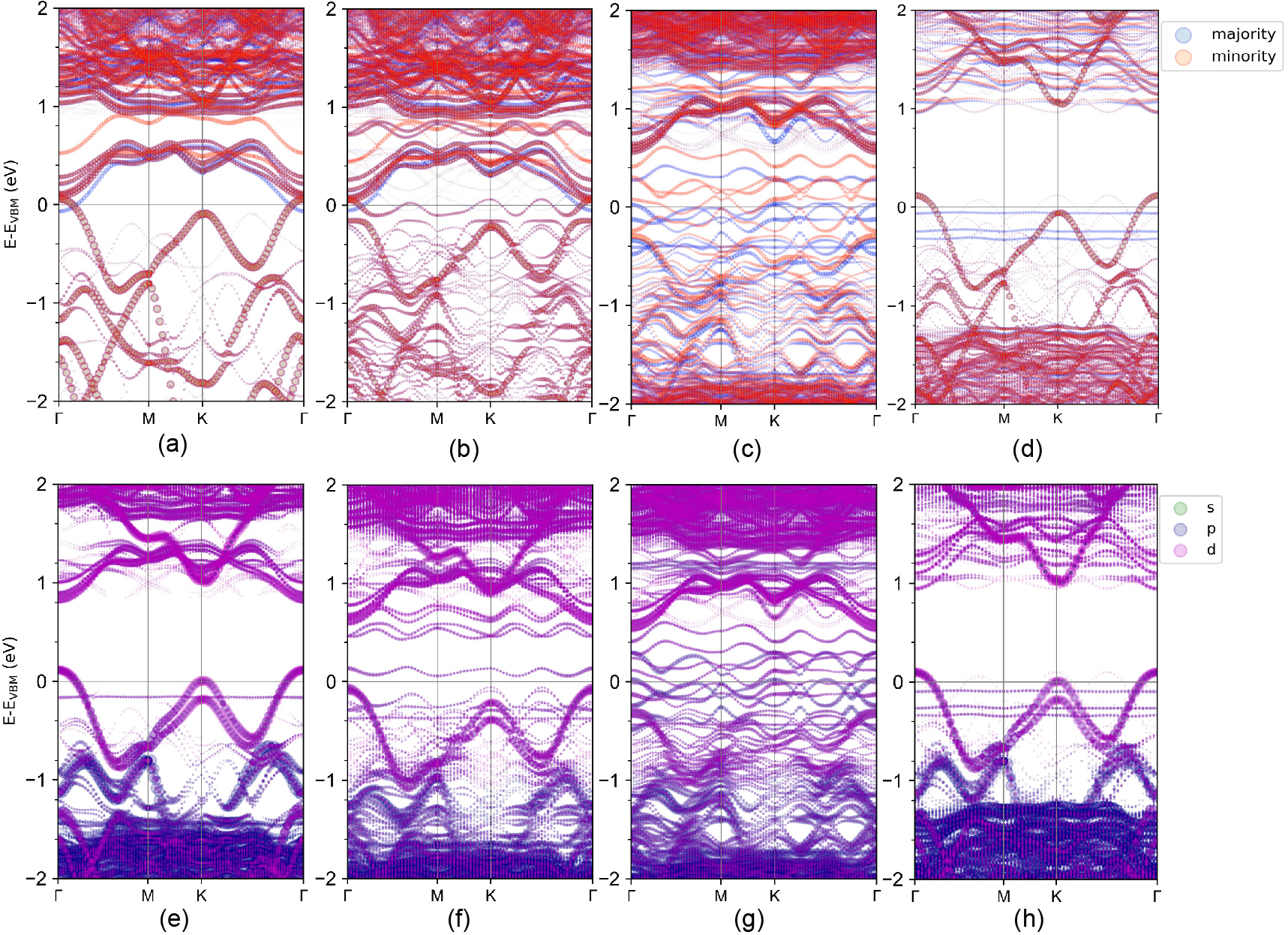}
    \caption{Spin polarized-collinear calculation of the band structures of (a) \ch{MoSe2/V2O5}, (b) \ch{(MoSe2}+\ch{V_{Se})/V2O5}, (c) \ch{(MoSe2}+\ch{V_{Mo})/V2O5} and (d) \ch{MoSe2/(V2O5}+\ch{V_{O})}. Non-collinear calculation of the band structures of (e) \ch{MoSe2/V2O5}, (f) \ch{(MoSe2}+\ch{V_{Se})/V2O5}, (g) \ch{(MoSe2}+\ch{V_{Mo})/V2O5} and (h) \ch{MoSe2/(V2O5}+\ch{V_{O})}.}
    \label{fig:band_SOC}
    \vspace{1mm}
    \hrule 
\end{figure}

The spin-orbit coupling (SOC) is a relativistic correction that has important consequences for the electronic properties of TMDs and their heterostructures. Here, we calculate the orbitally projected band structures of the considered systems by taking into account the SOC effect. The results are given in Fig. \ref{fig:band_SOC} together with the band structures for the majority and minority carriers when SOC is not taken into account.

